\documentclass[manuscript]{aastex}
\usepackage[bookmarks,bookmarksnumbered,colorlinks=true, citecolor=blue, linkcolor=black,breaklinks]{hyperref}
\usepackage{bm}
\usepackage{verbatim, amsmath,amssymb,amsthm}

\newcommand{\V}[1]{{\bm{\mathbf{\MakeLowercase{#1}}}}} 
\newcommand{\M}[1]{{\bm{\mathbf{\MakeUppercase{#1}}}}} 
\newcommand{\argmin}[1]{\underset{#1}{\operatorname{argmin}}\text{ }}

\shorttitle{A new multiband period estimation algorithm}
\shortauthors{Mondrik, Long, and Marshall}

\begin{document}

\title{A Multiband Generalization of the Analysis of Variance Period Estimation Algorithm and the Effect of Inter-band Observing Cadence on Period Recovery Rate}


\author{N. Mondrik\altaffilmark{1}}
\affil{George P. and Cynthia Woods Mitchell Institute for Fundamental Physics and Astronomy, and Department of Physics and Astronomy \\
Texas A \& M University, College Station, TX 77843-4242}
\altaffiltext{1}{Department of Physics, Harvard University, Cambridge, MA 02138}
\email{nmondrik@g.harvard.edu}

\author{J. P. Long}
\affil{Department of Statistics\\
Texas A \& M University, 3143 TAMU, College Station, TX 77843-3143}

\and

\author{J. L. Marshall}
\affil{George P. and Cynthia Woods Mitchell Institute for Fundamental Physics and Astronomy, and Department of Physics and Astronomy \\
Texas A \& M University, College Station, TX 77843-4242}


\begin{abstract}
We present a new method of extending the single band Analysis of Variance period estimation algorithm to multiple bands. We use SDSS Stripe 82 RR Lyrae to show that in the case of low number of observations per band and non-simultaneous observations, improvements in period recovery rates of up to $\approx$60\% are observed. We also investigate the effect of inter-band observing cadence on period recovery rates. We find that using non-simultaneous observation times between bands is ideal for the multiband method, and using simultaneous multiband data is only marginally better than using single band data. These results will be particularly useful in planning observing cadences for wide-field astronomical imaging surveys such as LSST. They also have the potential to improve the extraction of transient data from surveys with few ($\lesssim 30$) observations per band across several bands, such as the Dark Energy Survey.
\end{abstract}





\section{Introduction}

The period-luminosity relationship of variable stars, first discovered by Henrietta Leavitt and calibrated by Ejnar Hertzsprung \citep{her19}, is an important step in the astronomical distance ladder. With applications to measuring the Hubble constant \citep{rie2011} and mapping out Galactic substructure \citep{ses2010}, periodic variables are key science drivers for next-generation astronomical imaging surveys such as the Large Synoptic Survey Telescope (LSST, \citealt{ive2008}) and Gaia \citep{eye2015}. Approximately 50 million variable stars will be detected by LSST \citep{ses2007} and 18 million variables by Gaia \citep{eye2000}, therefore, automated classifiers must be relied upon to find the variable sources and determine the period of the source, if it is periodic.

Numerous period finding algorithms have been implemented over the years (see \citealt{gra2013} for a comparison of various algorithms). One common characteristic of most modern period finding algorithms is the use of observational data in a single band. For current generation transient surveys such as the intermediate Palomar Transient Factory (iPTF; 10-5000 observations in $R$ band for certain fields\footnote{http://www.ptf.caltech.edu/page/first\_data\_release}, \citealt{Law2009a}) and Optical Gravitational Lensing Experiment (OGLE; 400-500 observations in $I$ band for LMC objects e.g., \citealt{Sos2009a,Sos2009b,Uda1992}), the volume of data in any one band is sufficient to accurately determine the period, rendering the use of additional bands redundant. However, in multiband surveys in which only a limited number of observations are available in each band, single band algorithms can struggle due to poor phase coverage \citep{gra2013}.

Multiband period finding methods have been explored before, but the proposed methods require either simultaneous measurements \citep{sue2012} or require that a period be correctly recovered by a single band algorithm in the majority of bands sampled \citep{olu2012}. The former case puts strict requirements on observing strategy, while the latter still suffers from the inability of single band algorithms to return accurate results with limited observations. Only recently \citep{lon2014,vdp2015} have methods been proposed that are general in the sense of allowing arbitrary observation times and fully incorporating data from multiple bands into an algorithm. 

In this Letter, we propose a method to extend the Analysis of Variance (AoV, \citealt{sch1996}) single band algorithm to multiple bands. The method improves period recovery rates for poorly sampled multiband light curves. In addition, we discuss the importance of observational cadence between the bands to be used, and show that non-simultaneous observations between bands increases the ability of our multiband algorithm to recover the correct period.

\section{Data}
\label{sec:data}
In this Letter, we select a sample from the 483 RR Lyrae stars from \cite{ses2010}, and use light curves from the Sloan Digital Sky Survey (SDSS) Stripe 82 Variable Source Catalog \citep{ive2007}. These stars have a reasonably large number of observations, with a median number of observations per band of 56 across the SDSS \emph{g}, \emph{r}, and \emph{i} bands, and 55 in \emph{u} and \emph{z}. The data span 3340 days. Of the 483 sources found in \citet{ses2010}, 33 were either not found in the Variable Source Catalog, or had $<$10 observations in one or more bands. Table~\ref{tab:data} gives a complete description of the number of RR Lyrae as a function of number of observations and downsampling method (described in section~\ref{sec:testing}).  The three band sample uses only the $g$, $r$, and $i$ bands. It should also be noted that the typical time for SDSS to complete one pass through all filters is $\approx 0.004$ days (5.7 mins, \citealt{yor2000}).


\section{Method}
\subsection{AoV Multiband Extension}
\label{sec:MBext}
As our model for the variation of brightness with time in a single band, we adopt a sinusiod with $K$ harmonics.  Assuming $n_b$ observations in each of $B$ bands, our data are of the form $\{(t_{bi},m_{bi},\sigma_{bi})_{i=1}^{n_b}\}_{b=1}^B$, where $t_{bi}$ is the time of the $i$-th observation in band $b$, $m_{bi}$ is the measured magnitude at that time, and $\sigma_{bi}$ is the uncertainty associated with $m_{bi}$. We assume $\omega$, the frequency, is constant across all bands. Our model can be written as 
\begin{align}
\label{eq:model}
m_{bi} &= \beta_{0b} + \sum_{k=1}^K a_{bk} \sin(k\omega t_{bi} + \phi_{bk}) + \epsilon_{bi}\\
&= \beta_{0b} + \sum_{k=1}^K (a_{bk}\cos(\phi_{bk})\sin(k\omega t_{bi})  + a_{bk}\sin(\phi_{bk})\cos(k\omega t_{bi})) + \epsilon_{bi} \nonumber
\end{align}
where $\epsilon_{bi} \sim N(0,\sigma_{bi}^2)$ are independent across $i$ and $b$.  This model is equivalent to the  multiphase $N_{base}=0$, $N_{band}=K$ model of \citet{vdp2015}.  The periodogram we construct (see Equation \ref{eq:periodogram}, this article) is different than that of \citet{vdp2015}.  See Equation \ref{eq:periodogram} and the discussion in section \ref{sec:VI15} for more details. \citet{lon2014} studies this model with $K=1$ and termed it MGLS.  The authors did not construct periodograms for this model, and did not study the effects of inter-band observing cadence on period recovery.

One natural approach for estimating $\omega$ is to use maximum likelihood. Let $\V{a}_b = (a_{b1},\ldots,a_{bK})$ and $\V{a} = (\V{a_1},\ldots \V{a}_B)$. Analogous definitions apply for $\V{\phi}_b$ and $\V{\phi}$. Let $\V{\beta}_0 = (\beta_{01},\ldots, \beta_{0B})$. Since the error model is normal, maximum likelihood is equivalent to finding the $\omega$ which minimizes the weighted sum of squares, sometimes known as ``chi--squared minimization.''
\begin{align*}
\widehat{\omega} &= \argmin{\omega} \min_{\V{a},\V{\phi},\V{\beta_0}} \sum_{b=1}^B \sum_{i=1}^{n_b} \left(\frac{m_{bi} - \sum (a_{bk}\cos(\phi_{bk})\sin(k\omega t_{bi})  + a_{bk}\sin(\phi_{bk})\cos(k\omega t_{bi})) - \beta_{0b}}{\sigma_{bi}}\right)^2\\
&= \argmin{\omega}  \sum_{b=1}^B \min_{\V{a}_b,\V{\phi}_b,\beta_{0b}} \sum_{i=1}^{n_b} \left(\frac{m_{bi} - \sum (a_{bk}\cos(\phi_{bk})\sin(k\omega t_{bi})  + a_{bk}\sin(\phi_{bk})\cos(k\omega t_{bi})) - \beta_{0b}}{\sigma_{bi}}\right)^2.
\end{align*}
We moved the min inside the sum over $b$ because the $b^{th}$ summand only depends on $\V{a}_b,\V{\phi}_b,\beta_{0b}$. 

The sum over $i$ can be simplified by noting the linearity of the model and reparameterizing. Let $\V{m}_b = (m_{b1},\ldots,m_{bn_b})^T$. Let $\beta_{bk1} = a_{bk}\cos(\phi_{bk})$ and $\beta_{bk2} = a_{bk}\sin(\phi_{bk})$. Define $\V{\beta_b} = (\beta_{0b},\beta_{b11},\beta_{b12},\ldots,\beta_{bK1},\beta_{bK2})^T \in \mathbb{R}^{2K+1}$. Let $\M{\Sigma}_b$ be a $n_b \times n_b$ diagonal matrix where $\Sigma_{bii} = \sigma_{bi}^2$. Define
\begin{equation*}
\M{X}_b(\omega) = \begin{pmatrix}
1 & \sin(\omega t_{b1}) & \cos(\omega t_{b1}) & \dots  & \sin(K\omega t_{b1}) & \cos(K\omega t_{b1}) \\
1 & \sin(\omega t_{b2}) & \cos(\omega t_{b2}) & \dots  & \sin(K\omega t_{b2}) & \cos(K\omega t_{b2}) \\
\vdots  & \vdots  &\vdots  &\ddots & \vdots & \vdots  \\
1 & \sin(\omega t_{bn_b}) & \cos(\omega t_{bn_b}) & \dots  & \sin(K\omega t_{bn_b}) & \cos(K\omega t_{bn_b}) \\
\end{pmatrix} \in \mathbb{R}^{n_b \times (2K+1)}
\end{equation*}

We rewrite the ML estimator as
\begin{equation*}
	\widehat{\omega} = \argmin{\omega} \sum_{b=1}^B \min_{\V{\beta}_b}  \, (\V{m}_b - \M{X}_b(\omega)\V{\beta}_b)^T\M{\Sigma}^{-1}_b(\V{m}_b - \M{X}_b(\omega)\V{\beta}_b)
\end{equation*}
The problem is now identical to weighted least squares so the $\V{\beta}_b$ which minimizes the expression is 
\begin{equation*}
	\V{\widehat{\beta}}_b(\omega) = (\M{X}_{b}^T(\omega)\M{\Sigma}_b^{-1}\M{X}_b(\omega))^{-1}\M{X}_b(\omega)^T\M{\Sigma}_b^{-1}\V{m}_b
\end{equation*}
Define
\begin{equation}
	\label{eq:weighted}
	RSS_b(\omega) = (\V{m}_b - \M{X}_b(\omega)\V{\widehat{\beta}}_b(\omega))^T\M{\Sigma}^{-1}_b(\V{m}_b - \M{X}_b(\omega)\V{\widehat{\beta}}_b(\omega))
\end{equation}
and we have
\begin{equation}
	\label{eq:ml_est}
	\widehat{\omega} = \argmin{\omega} \sum_{b=1}^B RSS_b(\omega)
\end{equation}
One can reconstruct maximum likelihood estimators for the original parameterization from the $\V{\widehat{\beta}_b}(\widehat{\omega})$. From a computational perspective, the ML estimator requires performing $B$ weighted least squares estimates at each frequency.

Rather than obtain a single period estimate, it may be useful at any proposed $\omega$ to have a measure of the confidence that $\omega$ is the true frequency. Periodograms are functions which map frequencies to some measure of confidence. Often periodograms are constructed so that under the null hypothesis of no magnitude variation (i.e., $m_{bi} = \beta_{b0} + \epsilon_{bi}$), the periodogram has a known distribution at any particular frequency. We construct such a periodogram for the model specified by Equation \eqref{eq:model}. The frequency which maximizes this periodogram will be shown to be the maximum likelihood estimator in Equation \eqref{eq:ml_est}. This periodogram is a direct generalization of the AoV periodogram of \citet{sch1996} to multiband data because:
\begin{itemize}
	\item With a single band, the periodogram simplifies to the AoV periodogram of \citet{sch1996}.
	\item The periodogram retains the F distribution under the null hypothesis of constant magnitude in every band.
\end{itemize}

We now discuss how to construct the periodogram following the notation of Section \ref{sec:MBext}. We then go into further detail regarding the equivalence of this periodogram to \citet{sch1996}, and compare this periodogram to \citet{vdp2015}.

Under the notation of the previous section the model in Equation \eqref{eq:model} can be written as
\begin{equation}
	\label{eq:matrix_model}
	\V{m}_b = \M{X}_b(\omega)\V{\beta}_b + \V{\epsilon}_b
\end{equation}
where $\V{\epsilon}_b \sim N(0,\M{\Sigma}_b)$ for all $b$. Consider testing the null hypothesis
\begin{equation*}
	H_0: \V{m_b} = \V{1}\beta_{b0} + \V{\epsilon}_b \, \forall \, b.
\end{equation*}
This hypothesis states that the magnitude is a constant $\beta_{b0}$ in each band. Since the first column of $\M{X}_b(\omega)$ is $\V{1}$, this is a submodel of Equation \eqref{eq:matrix_model}. The weighted least squares estimator for the submodel is
\begin{equation*}
	\widehat{\beta}_{b0} = (\V{1}^T\M{\Sigma}_b^{-1}\V{1})^{-1}\V{1}^T\M{\Sigma}_b^{-1}\V{m}_b = \frac{1}{\sum_{i=1}^{n_b} \sigma_{bi}^{-2}}\sum_{i=1}^{n_b} \frac{m_{bi}}{\sigma_{bi}^2}
\end{equation*}
The residual sum of squares is
\begin{equation*}
	RSS^0_b = (\V{m}_b - \V{1}\widehat{\beta}_{b0})^T\M{\Sigma}_b^{-1}(\V{m}_b - \V{1}\widehat{\beta}_{b0})
\end{equation*}
Standard results in statistics (for example Sections 2.5 and 2.6 of \citealt{scheffe1959analysis}) show that under the null hypothesis
\begin{align*}
	RSS^0_b - RSS_b(\omega) \sim \chi^2_{2K}\\
	RSS_b(\omega) \sim \chi^2_{n_b-2K-1}
\end{align*}
where $\chi^2_j$ refers to a chi--squared distribution with $j$ degrees of freedom. Further these two quantities are independent. Since the sum of $\chi^2$ random variables is $\chi^2$ we have
\begin{align*}
	\sum_{b=1}^B RSS^0_b - \sum_{b=1}^B RSS_b(\omega) \sim \chi^2_{2KB}\\
	\sum_{b=1}^B RSS_b(\omega) \sim \chi^2_{\sum_{b=1}^B n_b-2KB-B}
\end{align*}
Finally we define the periodogram at frequency $\omega$ to be the ratio of these quantities divided by their respective degrees of freedom
\begin{equation}
	\label{eq:periodogram}
	\Theta(\omega) = \frac{(\sum_{b=1}^B n_b-2KB-B)\left(\sum_{b=1}^B RSS^0_b - \sum_{b=1}^B RSS_b(\omega)\right)}{2KB (\sum_{b=1}^B RSS_b(\omega))}
\end{equation}
Under $H_0$, $\Theta(\omega) \sim F_{2KB,\sum_{b=1}^B n_b-2KB-B}$. A few comments on the periodogram:
\begin{itemize}
	\item In practice, the frequency which maximizes the periodogram is often used as a period estimate. The frequency which maximizes the periodogram will minimize $\sum_{b=1}^B RSS_b(\omega)$, which we showed in Equation \eqref{eq:ml_est} is the maximum likelihood estimator.
	\item With a single band the periodogram becomes
	\begin{equation*}
		\Theta(\omega) = \frac{(n-2K-1)\left(RSS^0 - RSS(\omega)\right)}{2K (RSS(\omega))}
	\end{equation*}
	which matches the periodogram of \citet{sch1996} Equation 11 (although with different notation).
\end{itemize}

In addition to developing the single band AoV algorithm, \citet{sch1996} also developed a fast routine for evaluating $RSS_b(\omega)$ based on finding orthogonal polynomials on the unit circle.  For this reason, we use a Fortran implementation of the single band AoV algorithm\footnote{http://users.camk.edu.pl/alex/} as the basis for constructing our multiband periodogram. A small python code demonstrating how to do this been made available\footnote{https://github.com/Mondrik/Multiband\_AoV\_Demo}. As input parameters to the single band AoV algorithm used in this work, we use a minimum frequency of 1 day$^{-1}$, an upper frequency of 5 day$^{-1}$, a frequency step of 0.0001 day$^{-1}$ and one harmonic (corresponding to FR0=1, FRU=5, FRS=0.0001, and NH2=2 in the AoV code). 

This periodogram is different from that of \citet{vdp2015} (see Equation 22 in their paper).  They use $RSS_b^0$ in the denominator instead of $RSS_b(\omega)$.  The unregularized models of \citet{vdp2015} follow an incomplete beta distribution (see \citet{sch1998}, Eqn. 6). It should also be noted that this multiband method implicitly assumes that the period of oscillation is the same for each band.  If the period varies significantly across bands, this method will not be suitable for use.

\begin{figure*}
	\figurenum{1}
	\label{fig:multi}
	\plotone{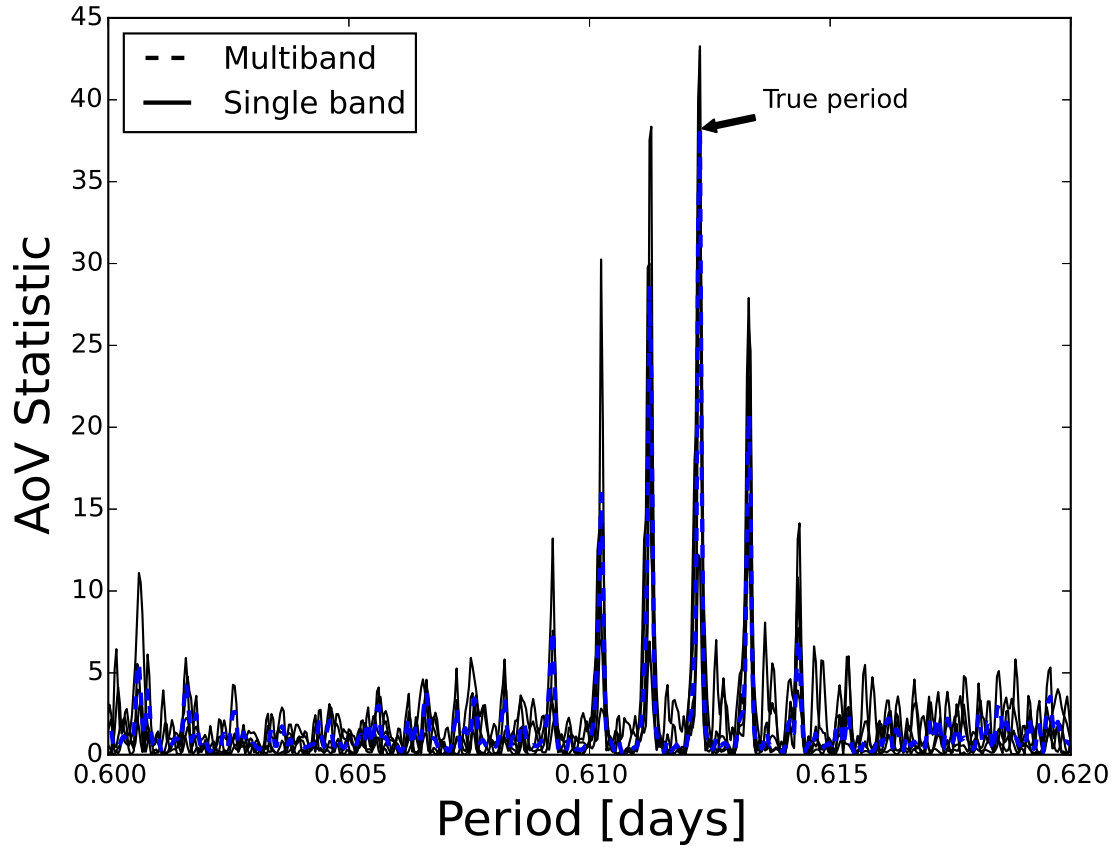}
	\caption{The single band AoV periodograms and multiband periodogram constructed using Equation~(\ref{eq:periodogram}). The AoV statistic is an indication of how well the trial function of the AoV algorithm fits the light curve folded with period $P$, with higher values indicating a better fit. The multiband periodogram is given by the blue dashed line, while the \emph{ugriz} single band periodograms are given by the solid black lines. The periodogram was generated using non-simultaneous downsampling (to 19 observations per band in 5 bands) of an RR Lyrae from \citet{ses2010}.}
\end{figure*}

Figure~\ref{fig:multi} compares the multiband periodogram with its single band components.


\subsection{Testing the Algorithm}
\label{sec:testing}
\begin{figure*}
	\figurenum{2}
	\label{fig:comp}
	\epsscale{0.9}
	\plotone{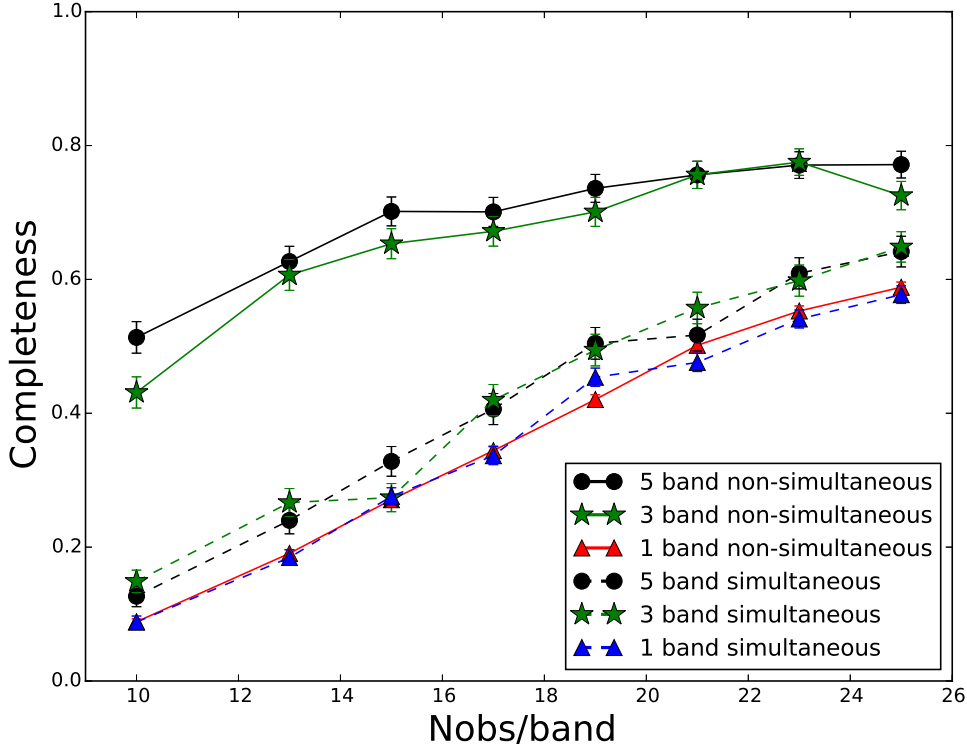}
	\epsscale{1}
	\caption{Completeness fraction (number of periods correctly recovered divided by number of light curves attempted, period is successfully recovered if $|P_{Alg} - P_{True}| \leq 0.001$ days.) as a function of observations per band across the SDSS \emph{ugriz} bands. The 483 RR Lyrae of \cite{ses2010} were used as a dataset, as described in section~\ref{sec:data}. Single band periods and one multiband period were estimated for each object. The single band completeness is obtained by dividing the number of correct single band period identifications by the total number of single band light curves attempted (i.e., 5x, 3x, or 1x the number of RR Lyrae), while multiband completeness is calculated by dividing the number of multiband correct period identifications by the number of RR Lyrae attempted. The solid (dashed) black line with circles represent the completeness fraction for 5 band (\emph{ugriz}) non-simultaneous (simultaneous) downsampling. The green solid (dashed) lines and star markers is the same, but for 3 bands (\emph{gri}) of data. The single band counterparts for non-simultaneous and simultaneous downsampling are given by the solid red line and dashed blue line with triangles, respectively. For simultaneous observations, the typical separation between observations is $\lesssim 0.005$ days.} 
\end{figure*}

To test the algorithm, we downsample the number of observations per band for each light curve using both simultaneous and non-simultaneous downsampling. For non-simultaneous downsampling, observations (consisting of a time of observation, band, magnitude, and photometric error) are randomly selected from all available observations. Observations are selected until all bands have $n_{obs}$ observations. For simultaneous downsampling, an observation in one band is chosen. We then choose observations in the other bands such that the absolute difference in observation times is $\leq 0.005$ days (7.2 minutes) from the initial observation time. This is repeated $n_{obs}$ times. Since the time for SDSS to complete one pass through all filters is about 5.7 minutes \citep{yor2000}, these observations are as close in phase space as possible. We also choose to use a flat time difference rather than a fraction of the known period in order to mimic the lack of \emph{a priori} knowledge of the variable object, as is the case in survey planning. We then define a period as correctly recovered if $|P_{Alg} - P_{True}| \leq 0.001$ days, where $P_{Alg}$ is the period corresponding to the largest value of the multiband or single band AoV periodogram, and $P_{True}$ is the period as measured by \citet{ses2010}. The results are shown in Figure~\ref{fig:comp}.  Error bars are estimated by assuming a binomial distribution at each $n_{obs}$ characterized by the estimated completeness and number of objects.


\section{Discussion}
\subsection{Benefit of Multiband over Single Band}
The most striking result shown in Figure~\ref{fig:comp} is the large separation between the multiband non-simultaneous completeness fraction and the single band non-simultaneous completeness fraction; even the use of as few as 3 bands can significantly improve recovery rates over single band methods. For surveys with a low number of observations per band ($n_{obs} \lesssim$ 30), such as the Dark Energy Survey \citep{des2005}, multiband methods can provide a significant increase in fraction of correctly recovered periods, allowing for more accurate classification of transient objects in the survey.

\subsection{Impact of Inter-band Observing Cadence}
The second major result noticeable in Figure~\ref{fig:comp} is the difference between the simultaneous and non-simultaneous downsampling groups. It should be noted that in the single band case, simultaneous and non-simultaneous downsampling should have no effect, so the scatter between the two is indicative of the randomness in choosing the observations. The difference between multiband non-simultaneous and simultaneous downsampling arises primarily from the increase in phase space coverage of the non-simultaneous downsampling relative to the simultaneous downsampling. In the case of simultaneous downsampling, the additional bands add little new information about the light curve not contained in other bands, leading to poorer performance, despite having the same number of total observations.

\begin{figure*}
	\figurenum{3}
	\label{fig:per-per}
	\plottwo{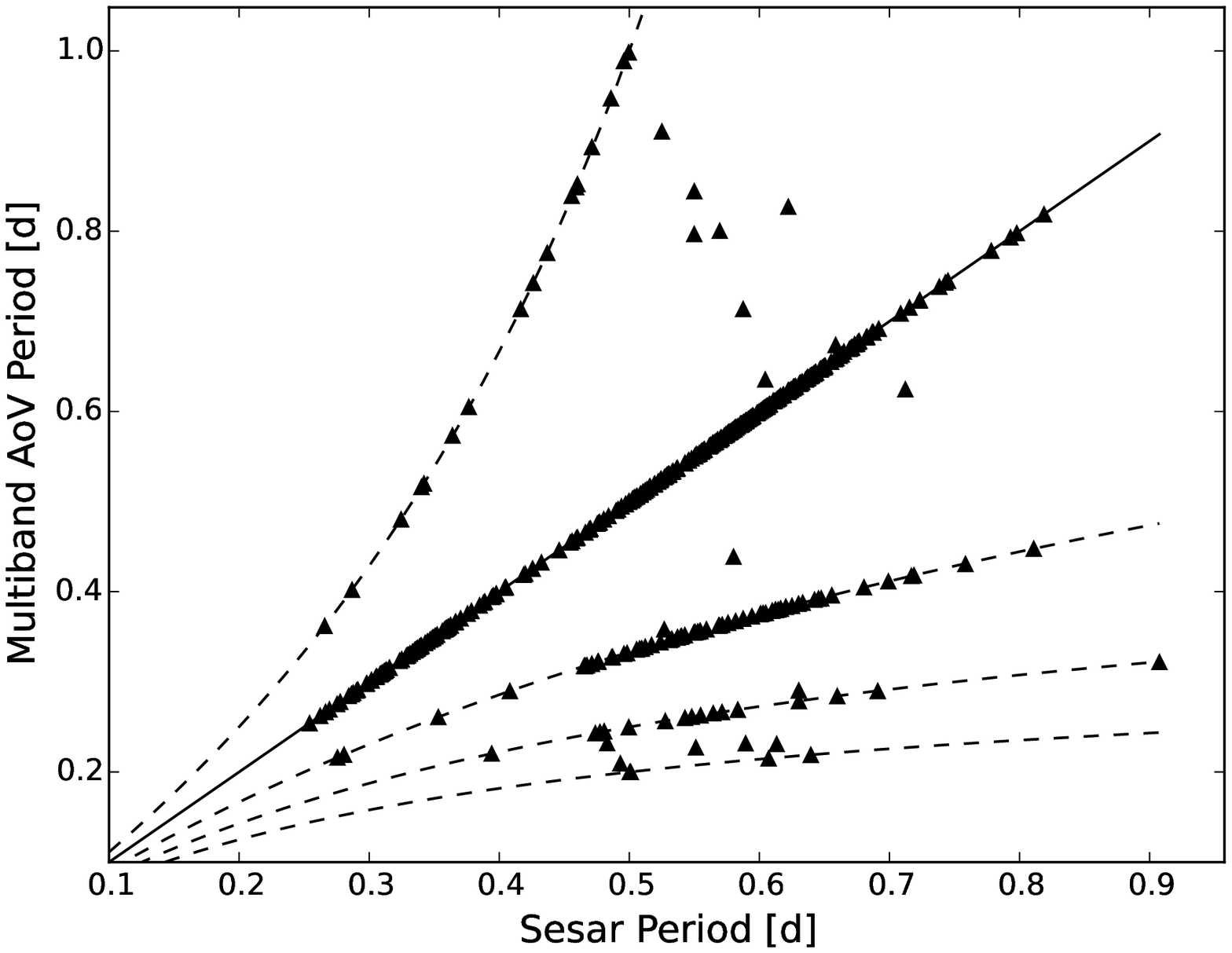}{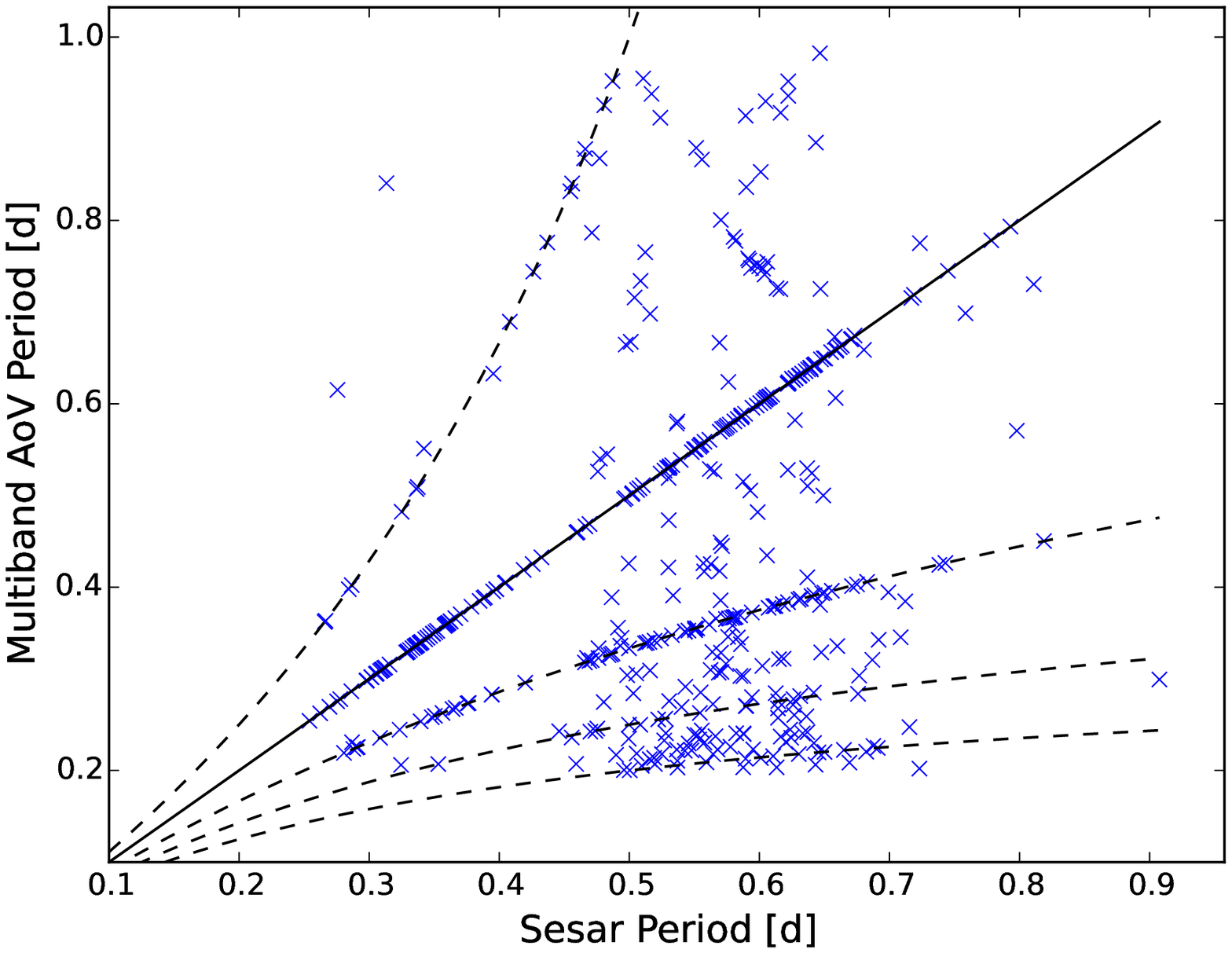}
	\caption{Comparison of best fit periods from this work to \citet{ses2010} for the multiband method in the non-simultaneous (left) and simultaneous (right) cases.  Periodograms were made using 15 observations per band. The solid line represents a 1:1 match, while the dashed lines represent beat frequencies.  In general, the non-simultaneous data set fails along beat frequencies, while the simultaneous data set fails in a more random manner.}
\end{figure*}

Figure~\ref{fig:per-per} demonstrates the failure modes of our multiband model in the case of non-simultaneous and simultaneous observations.  We plot the best fit period from the AoV multiband model with 15 observations per band against the period taken from \citet{ses2010}.  In the non-simultaneous case, the multiband method fails primarily along beat periods, given by $P_n = P/(1+nP)$ for integer $n$, as discussed in \citet{vdp2015}.  In the simultaneous case, the multiband method tends to fail in a much more random fashion.


\subsection{Constructing Data Sets for Maximized Period Recovery Rates}
The improvement of recovery rates with unique phase space observations (non-simultaneous observations) suggests that recovery rates cannot be significantly improved by downsampling 25 simultaneous observations to, for example, sets of 12 observations that are less simultaneous. Fundamentally, there are only 25 unique phase space observations of the object, which constrains the maximum amount of phase space separation between the bands. In order to construct a truly non-simultaneous dataset, we require a minimum of $n_{obs}$ per band times the number of bands. Hence our non-simultaneous downsampling set is not truly non-simultaneous, since we are randomly downsampling from less than 125 observations across 5 bands (at 25 observations per band). Since the typical number of observations per band is $\approx 55$, it is impossible for us to separate the observations completely for $n_{obs}>11$ in 5 bands. In this case, the algorithm is limited by the construction of the data set, i.e., how non-simultaneous the observations are. It would therefore be advantageous to construct a data set that is as non-simultaneous as possible, rather than downsample from a simultaneous data set, in order to use the maximum number of observations.

\subsection{Implications for Future Imaging Surveys}
This method of constructing a data set has a major potential impact on observational cadence planning for upcoming wide-field imaging surveys such as LSST. By varying observation times between bands, accurate periods for variable sources can be estimated much sooner than otherwise possible. The phase space effect also has implications for our ability to extract transient data from ongoing surveys such as the Dark Energy Survey, which could see a boost in period recovery rates by employing an algorithm similar to our proposed method.

\section{Comparison with VanderPlas \& Ivezi{\'c} (2015)}
\label{sec:VI15}
As we mentioned earlier, another method similar to ours in spirit is that of \citet{vdp2015}.  Both methods use truncated Fourier series to model given observations across an arbitrary number of bands.  However, the method used in \citet{vdp2015} is effectively an extension of the Lomb-Scargle periodogram \citep{sca1982,lom1976}, while ours extends the Analysis of Variance periodogram.  In the single band case, \cite{gra2013} show that the multiharmonic AoV algorithm (AOVMHW in their notation, AoV in ours) tends to be among the top performers in any test of period estimation.  \citet{sch1998} asserts that statistically, the use of PDM \citep{jur1971,ste1978}, AoV, and $\chi^2$ statistics is largely a matter of taste,   although it will be interesting to perform another analysis similar to that of \citet{gra2013} on multiband data using methods such as ours and that of \citet{vdp2015} to determine when each algorithm is most effective.

\section{Conclusion}
We have introduced a new method of estimating periods of periodic variables using multiband imaging data. We extended the existing AoV period estimation algorithm to incorporate data from multiple bands while maintaining the fundamental characteristics of the single band algorithm. This allows for the use of relatively few observations ($\approx25$) per band across several bands while maintaining a reasonable level of completeness ($\approx 70-80\%$). We have also shown the importance of the (non-)simultaneity of observation timing. For a fixed number of observations per band, non-simultaneous observations offer better opportunity for period recovery than simultaneous observations. This effect of observational simultaneity has implications for the area of survey planning, particularly in the early period of surveys such as LSST, when the volume of data is not enough to render multiband period estimation redundant. It also has implications for non-transient surveys imaging fields at multiple epochs. By carefully choosing the observation time and band, our proposed mutiband algorithm can extract periods from data previously considered too poorly sampled to be of use.

\section{Acknowledgements}
The authors would like to thank Ting Li for her comments regarding data set construction. This work was supported by NSF grant AST-1263034, ``REU Site: Astronomical Research and Instrumentation at Texas A\&M University''. Funding for the SDSS and SDSS-II has been provided by the Alfred P. Sloan Foundation, the Participating Institutions, the National Science Foundation, the U.S. Department of Energy, the National Aeronautics and Space Administration, the Japanese Monbukagakusho, the Max Planck Society, and the Higher Education Funding Council for England. The SDSS Web site is http://www.sdss.org/.

\bibliography{adssample}

\begin{deluxetable}{ccccc}
	\tabletypesize{\scriptsize}
	\tablecolumns{5}
	\tablecaption{Number of RR Lyrae in sample\label{tab:data}}
	\tablehead{\colhead{$n_{obs}$ per band} & \colhead{Non-simultaneous 5 band} & \colhead{Simultaneous 5 band} & \colhead{Non-simultaneous 3 band} & \colhead{Simultaneous 3 band}}
	\startdata
	10 & 450 & 450 & 450 & 450 \\
	13 & 450 & 449 & 450 & 450 \\
	15 & 450 & 448 & 450 & 449 \\
	17 & 448 & 447 & 448 & 448 \\
	19 & 448 & 445 & 448 & 447 \\
	21 & 447 & 445 & 447 & 445 \\
	23 & 445 & 440 & 445 & 443 \\
	25 & 444 & 438 & 444 & 441 \\
	\enddata
\end{deluxetable}

\end{document}